\documentclass[aps,prl,twocolumn,superscriptaddress]{revtex4-1}
\usepackage{graphicx} 
\usepackage{epsfig}

\begin{document}


\title{The Suppression of Radiation Reaction and Laser Field Depletion in Laser-Electron beam interaction}


\author{J. F. Ong}
\affiliation{National Institute for Physics and Nuclear Engineering, ELI-NP, Str Reactorului, nr. 30, P.O.Box MG-6, Bucharest-Magurele 077125, Romania.}
\affiliation{Dept. of Physics, Graduate School of Science, Osaka University
1-1 Machikaneyama, Toyonaka, Osaka 560-0043, Japan}
\affiliation{Research Center for Nuclear Physics, 10-1 Mihogaoka Ibaraki, 567-0047 Osaka, Japan}
\author{T. Moritaka}
\affiliation{Fundamental Physics Simulation Research Division, National Institute for Fusion Science,
322-6 Oroshi-cho, Toki City, GIFU Prefecture, 509-5292 Japan}
\author{H. Takabe}
\affiliation{Helmholtz-Zentrum Dresden-Rossendorf (HZDR), Bautzner Landstrasse 400, 01328 Dresden, Germany}

\date{\today}
\begin{abstract}
The effects of radiation reaction (RR) have been studied extensively by using the ultraintense laser interacts with the counter-propagating relativistic electron. At the laser intensity at the order of $10^{23}$ W/cm$^2$, the effects of RR are significant in a few laser period for a relativistic electron. However, the laser at such intensity is tightly focused and the laser energy is usually assumed to be fixed. Then, the signal of RR and energy conservation cannot be guaranteed. To assess the effects of RR in a tightly focused laser pulse and the evolution of the laser energy, we simulate this interaction with a beam of $10^9$ electrons by means of Particle-in-Cell (PIC) method. We observed that the effects of RR are suppressed due to the ponderomotive force and accompanied by a non-negligible amount of laser field energy reduction. This is due to the ponderomotive force that prevents the electrons from approaching the center of the laser pulse and leads to the interaction at weaker field region. At the same time, the laser energy is absorbed through ponderomotive acceleration. Thus, the kinetic energy of the electron beam has to be carefully selected such that the effects of RR become obvious.
\end{abstract}


\maketitle

\section{Introduction\label{Introduction}}

The numerical studies of gamma-ray production by including the effects of radiation reaction (RR) have been performed extensively \cite{PhysRevX.2.041004,PhysRevLett.108.195001,MyPoP}. In which it was mentioned that the cumulative effects of RR in classical radiation-dominated regime become strong at the laser intensity  $10^{22}$ W/cm$^2$ with many cycle laser pulse \cite{MyPoP}. Hence, the cumulative effects of RR are certainly strong at laser intensity $10^{23}$ W/cm$^2$ even with a few-cycle laser pulse only. This provides the feasibility to observe the effects of RR in the near-future experimental facility such as Extreme Light Infrastructure (ELI) \cite{ELI}. However, at such facility, the laser have to be tightly focused into a waist radius a few order of laser wavelength to achieve the mentioned laser intensity. Then, the treatment of the laser field is necessary beyond the paraxial approximation \cite{McDonald,McDonald2,parax}. Accordingly, the ponderomotive force comes into play when the finite spatial profile of a  laser fields is considered. A relativistic electron interacts with such field is reflected away from the center of laser field when the kinetic energy is less than the ponderomotive potential \cite{PhysRev.150.1060,Narozhny2000}. Such effects alert the expected outcome in the effort of spotting the effects of RR as compared to the plane wave field.

On the other hand, an optimal conversion efficiency was observed by colliding an electron of the energy of 40 MeV with a counter-propagating laser field at the intensity of $2\times10^{23}$ W/cm$^2$ with 10 fs of pulse duration \cite{MyPoP}. If one notes on the laser pulse energies, they were assumed to be constant and the fields were not solved by Maxwell's equations. However, Ref. \cite{MyPoP} reported that the electromagnetic (EM) work was enhanced at laser intensity above $10^{22}$ W/cm$^2$ when the effects of RR are taken into account. This is only for the case of a single particle not to mention when a bunch of $10^9$ electrons involved in the interaction. Thus, the laser energy can no longer be assumed to be fixed because the energy conversion efficiency will be overestimated. Besides, the laser energy cannot be defined for a 2-dimensional system (2D) due to the absence of the spatial variation at the laser focusing plane. To study the laser field energy variation, a 3-dimensional (3D) system is required. 

As a solution to the aforementioned issues, we solve Maxwell's equations by taking into account the current density generated by the motion of the electron beam by using 3D Particle-in-Cell (PIC) method. Thus, the energy conversion among laser, electron beam and radiation can be estimated consistently. This paper reports the suppression of the effects of RR in the tightly focused laser pulse and identifies a non-negligible amount of laser field energy depletion.

In the following sections, we first discuss the applicability of the Sokolov's model such that the concept of proper time is approximately validated. We then compare the magnitude of the RR force to the ponderomotive force and then the energy balance is derived. The results for an electron beam with the energy of 40 MeV and 1 GeV interacting with a tightly focused laser pulse are presented. 

\section{Radiation Reaction\label{Sokolov} }
The equation of motion including the effects of RR discussed in Ref. \cite{SokolovMourou, SokolovModeling, SokolovJETP} do not preserve the identity $\dot{x}^{2}=c^2 $, where $\dot{x}$ is the four-velocity and $c$ is the speed of light. The velocity of an accelerated electron contains an additional term. In the frame where the electron momentum, $\mathbf{p}=$ 0, the velocity of the electron does not vanish. In this situation, the additional term to the velocity is written as $\mathbf{\dot{x}}=\tau_0 e\mathbf{E}/m$, where $e$ and $m$ are the charge and mass of the electron respectively, $\tau_0=2e^2/3mc^3$, and $\mathbf{E}$ is the external electric field. The non-vanishing velocity is interpreted as an extra work done by the external field to compensate the energy loss. Therefore, it leads to the impossibility to have an instantaneous rest frame for a radiating particle and the concept of proper time changed \cite{Sokolov2}.

As a result, in Sokolov's model, the identity $\dot{x}^{2}=c^2 $ is violated by a term proportional to $\chi^2$:
\begin{eqnarray}
\dot{x}^{2}=c^2\left(1-\frac{\chi^{2}}{137^{2}} \right)\label{x^2}
\end{eqnarray} 
where $\chi$ is a quantum parameters. For an electron counter-propagates with respect to the laser field, the quantum parameter $ \chi=2(\hbar \omega_0 /mc^2)\gamma_0 a_0$, where $\hbar$ is the reduced Planck's constant, $\omega_0$ is the laser angular frequency, $\gamma_0$ is the initial Lorentz factor, and $a_0$ is the normalized laser amplitude. 

In order to keep the concept of proper time in a numerical simulation, then Sokolov's equation should be subject to the applicability limit so that $\dot{x}^{2}\approx c^2 $ is approximately preserved. The peak limit of the applicability can then be written as
\begin{eqnarray}
\frac{10^{-6}}{\lambda [\mu \text{m}]}\gamma_0 a_0 <<1\label{inequality}
\end{eqnarray}
where $\lambda [\mu \text{m}]$ is the laser wavelength in the unit of a micrometer. The identity is preserved within $ 0.1\%$ for $\chi<4$. Then, the quantum effects such as electron recoil is also considered. If the parameters of electron and laser do not satisfy the inequality (\ref{inequality}) then the concept of proper time has to be redefined. The near-future facility such as ELI-NP \cite{ELINP} equipped with two 10 PW lasers and a linac of 700 MeV electron beam would be able to provide the opportunity to test Eq. \ref{x^2}.

As the laser intensity reaches the order of $10^{23}$ W/cm$^{2}$, the effects of RR in one laser cycle would be significant for a relativistic electron counter-propagate with respect to the laser pulse. In addition, these effects are enhanced when the electron propagates through many-cycle laser field so that the cumulative effects of RR become strong, such that $R_CN\sim1$ where $R_C=2/3\alpha a_0\chi$ is a measurement of the strength of classical RR,  $\alpha=1/137$ is the fine structure constant and $N$ is the number of laser cycle \cite{MyPoP}. Nevertheless, the condition $R_CN\sim1$ may not be sufficient to determine the effects of RR due to the ponderomotive force when a laser field is focused into the waist of the order of several wavelengths. 


\section{Ponderomotive Force}\label{ponderomotive}
\noindent Apart from the effects of radiation reaction, the ponderomotive force also plays a crucial role in laser-electron beam collision. In the relativistic regime, the ponderomotive force is related to the effective mass of the electron in the field as $f_P=-\nabla m_{\text{eff}} c^2$ where $m_{\text{eff}}=m\sqrt{1+ <\mathbf{a}^2>}$, $ \mathbf{a}=e\mathbf{A}/mc^2$ and $\mathbf{A}$ is the vector potential of the external field \cite{macchi2013superintense}. The charged particle subject to this force tends to move into regions of diminishing field strength. However, a relativistic charged particle may overcome this limit and penetrate into strong field region \cite{PhysRev.150.1060,Narozhny2000}. The condition of penetrability was found to be $\gamma>\sqrt{1+a^2_0}$ for a focused laser field. In other words, the kinetic energy has to be larger than the potential energy, i.e. the ponderomotive potential. For $a_0>>1$, we can then write $\kappa>1$ for the condition of penetrability such that $\kappa=\gamma/a_0$. 

Moreover, an intensity threshold exists where the particles are not able to penetrate into the peak field strength region regardless their initial kinetic energy when the effects of RR are considered \cite{PhysRevA.90.053847}. Due to the effects of RR, an initially incoming particle with $\kappa>>1$ lost energy before reaching the core of the field and turn to the condition $\kappa<1$. The threshold of the impenetrability is $a_0\gtrsim a_{0*}(4\mu)^{1/3}$ where $a_{0*}=(3m/2e^2\omega)^{1/3}\simeq 470$ and $\mu=\lambda_0/(\pi w_0)$ is the diffraction angle and $w_0$ is the laser waist radius. For $w_0=2 ~ \mu\text{m}$ and $\lambda=1~\mu\text{m}$, the impenetrability threshold is  $a_0\gtrsim $ 400. 

Below the impenetrability threshold, we compare the magnitudes of the RR force to ponderomotive force in relativistic regime. The magnitude of the leading term of RR force is the order of $f_{RR}\sim 4\pi\alpha \hbar\omega_0\gamma^2a_0^2/(3\lambda)$ while the ponderomotive force is $f_P\sim-mc^2/(2\gamma)\nabla a^2$. By assuming a Gaussian profile at the focusing plane, then the vector potential is $\mathbf{a}=a_0\exp(-r^2/w_0^2)e^{i\eta}\hat{\mathbf{y}}$. Subsequently, the ratio can be written as
\begin{eqnarray}
	\frac{f_{RR}}{f_P}\sim \frac{2}{3}\pi\alpha\frac{\hbar\omega_0}{mc^2}\frac{w_0}{\lambda}\gamma^3 \label{frr/fp}
\end{eqnarray}
where $r^2=y^2+z^2$ and $\eta=\omega t -kx$. For the non-relativistic case, $\gamma\sim 1$, Eq. \ref{frr/fp} reduces to the one defined in Ref. \cite{PhysRev.150.1060}. 
 
The ponderomotive force is vanishing in the limit of infinite spatial extension (i.e. $w_0\rightarrow \infty$) for a charged particle with a given initial energy. Then, the effects of RR can be readily observed providing $R_CN\sim 1$. On the contrary, the effects of RR are suppressed by reducing the laser waist radius to a few orders of the laser wavelength even if $\gamma>>1$. Nevertheless, we limit the waist radius larger than the laser wavelength in order to validate the assumption for the ponderomotive force.

\section{Energy balance}
\noindent In PIC method, the fields evolve according to Maxwell's equations. During the interaction, the energy is transferred from field to particle or vice versa. Thus, it is important to check the energy balance for the processes that include the effects of RR. The energy equation for the field is written as 
\begin{eqnarray}
	\frac{ \partial W_{EM} }{ \partial t } + \nabla \cdot \mathbf{S} = - \mathbf{J} \cdot \mathbf{E} \label{ENEfield}
\end{eqnarray}
%
where $W_{EM}=\frac{1}{2}\left[ \epsilon_0\mathbf{E}^2 +\frac{1}{\mu_0}\mathbf{B}^2  \right]$ is the EM field total energy density, $\mathbf{S}$ is the Poynting vector, and $\mathbf{J}$ is the current density. Integrating over the volume on both side of Eq. \ref{ENEfield} and substitute the particle energy equation 
\begin{equation}
	\frac{d\mathcal{E}_{e}}{dt}= \int \mathbf{J} \cdot \mathbf{E}   d^3x- \frac{d\mathcal{E}_{rad}}{dt},\label{ENEpar}
\end{equation}
%
one obtains
\begin{equation}
	\frac{ \partial }{ \partial t } \left[\int W_{EM} d^3x + \mathcal{E}_{e} +    \mathcal{E}_{rad}              \right]=- \oint \mathbf{S}   \cdot d\mathbf{A}.\label{Poy}
\end{equation}
%
where $ \mathcal{E}_{rad} $ is the energy of emitted radiation as described in Ref. \cite{SokolovModeling}, and $\mathbf{A}$ is the surface area. Integrating both sides of this equation and assuming the interaction domain goes to infinity, i.e. the right hand side of Eq. \ref{Poy} goes to zero, we can write the energy balance equation as 
 \begin{equation}
	\frac{ \Delta\mathcal{E}_{field}  + \Delta\mathcal{E}_{e} +   \Delta \mathcal{E}_{rad} }{ E_{initial} }            =0\label{Etotal}
\end{equation}
where $ E_{initial}=E_{e}+E_{laser} $, $\Delta\mathcal{E}_{field}=\mathcal{E}_{field}(t)-\mathcal{E}_{field}(t=0)$, $\Delta\mathcal{E}_{e}=\mathcal{E}_{e}(t)-\mathcal{E}_{e}(t=0)$ and $\Delta\mathcal{E}_{rad}=\mathcal{E}_{rad}(t)-\mathcal{E}_{rad}(t=0)$. $\mathcal{E}_{rad}(t=0)$ is assumed to be zero. The change of the laser field energy is $\Delta\mathcal{E}_{field}$ while $\Delta\mathcal{E}_{e}$ and $\Delta \mathcal{E}_{rad}$ are the change of electron energy and the radiation emission energy respectively.

\section{Setting}
\noindent The fields of a tightly focused laser beam are described beyond paraxial approximation. Here, the pulsed Gaussian beam beyond paraxial approximation up to 5$^{\text{th}}$ order correction is considered \cite{McDonald,McDonald2,parax}. A possible choice for the temporal pulse function is $g(\eta)=1/\cosh (\eta/\eta_0)$ where $\eta=\omega t -kx$, $\eta_0=\omega_0t_L =2\pi N$, $t_L$ is the laser pulse duration, and $N$ is the number of laser cycle such that the condition $g'/g \ll 1$ is satisfied \cite{McDonald2}. A laser pulse of 10 fs pulse duration propagates in the $+ x$ direction with 2 $\mu$m waist radius is considered. For a laser pulse with $\lambda = 1$ $\mu \text{m}$ and the peak intensity of $2\times10^{23}$ W/cm$^{2}$ ($a_0 \sim 380$), inequality (\ref{inequality}) is satisfied for $\gamma_0<2000$. Thus, the initial energy of the electron beam are kept below 1 GeV.


Next, an electron beam is propagating in the opposite direction with the following phase space distribution:
\begin{equation}
    f(\mathbf{x},\mathbf{p})=\exp \left[    -\frac{x^2}{2\sigma_L^2} -\frac{y^2 + z^2 }{2\sigma_T^2}  -\frac{p_i^2}{2\sigma^2_{p_i}}   \right]
\end{equation}
where $p_i^2/\sigma^2_{p_i}=p_x^2/\sigma^2_{p_x} + p_y^2/\sigma^2_{p_y} + p_z^2/\sigma^2_{p_z}$,  $\sigma_{p_i}$ is the momentum spread while $\sigma_L$ and $\sigma_T$ being the longitudinal and transverse beam spread respectively. The values of $\sigma_L$ and $\sigma_T$ are both 1 $\mu$m so that all the electron in the beam interact with the laser pulse. Meanwhile, the momentum spread is 5 $\%$ of the initial momentum of the electron. 

The size of the simulation domain is $x \times y \times z= \text{30}\lambda\times\text{10}\lambda\times\text{10}\lambda$ with 1200 $\times$ 200 $\times$ 200 cells. There are a total of 40000 superparticles to represent $10^9$ electrons. The time step is taken to fulfil the Courant condition where $c\Delta t=0.5 \Delta x$. For a 40 MeV electron beam with $10^9$ electrons and at laser intensity of $2\times10^{23}$ W/cm$^2$, $E_{initial}=254.31$ J while $E_{initial}=254.46$ J for 1 GeV electron beam.

The collective behavior is weak as the Coulomb repulsive force is compensated by the magnetic force for two electrons travel in parallel at relativistic speed. This is a consequence of Lorentz transformation of the EM field around the relativistic electrons. Hence, only the EM fields of the laser can give a significant Lorentz force  on the electrons. 

Meanwhile, the emission spectra are obtained according to the method described in Ref. \cite{MyPoP}. The motion of the electron and the emission energy are computed according to the Sokolov's equations. This emission energy is then used to calculate the emission spectra. In the absence of RR, the motion of the electron is evaluated according to Lorentz force only and the emission energy is calculated according to the quantum-corrected relativistic Larmor formula.

The normalization of the physical variables are: $\mathbf{x}' \rightarrow \mathbf{x}/\lambda, t'\rightarrow ct/\lambda,\mathbf{p}'\rightarrow \mathbf{p}/mc,\mathbf{E}'\rightarrow e\mathbf{E}\lambda/mc^2,\mathbf{B}'\rightarrow e\mathbf{B}\lambda/mc, \mathbf{J}'\rightarrow 4\pi^2\mathbf{J}/cen_{cr}$, and $\rho'\rightarrow 4\pi^2\rho/en_{cr}$ where $n_{cr}$ is the critical density. 
\section{Results and Discussion}
\begin{figure}[h]
\begin{center}
\includegraphics[scale=0.075]{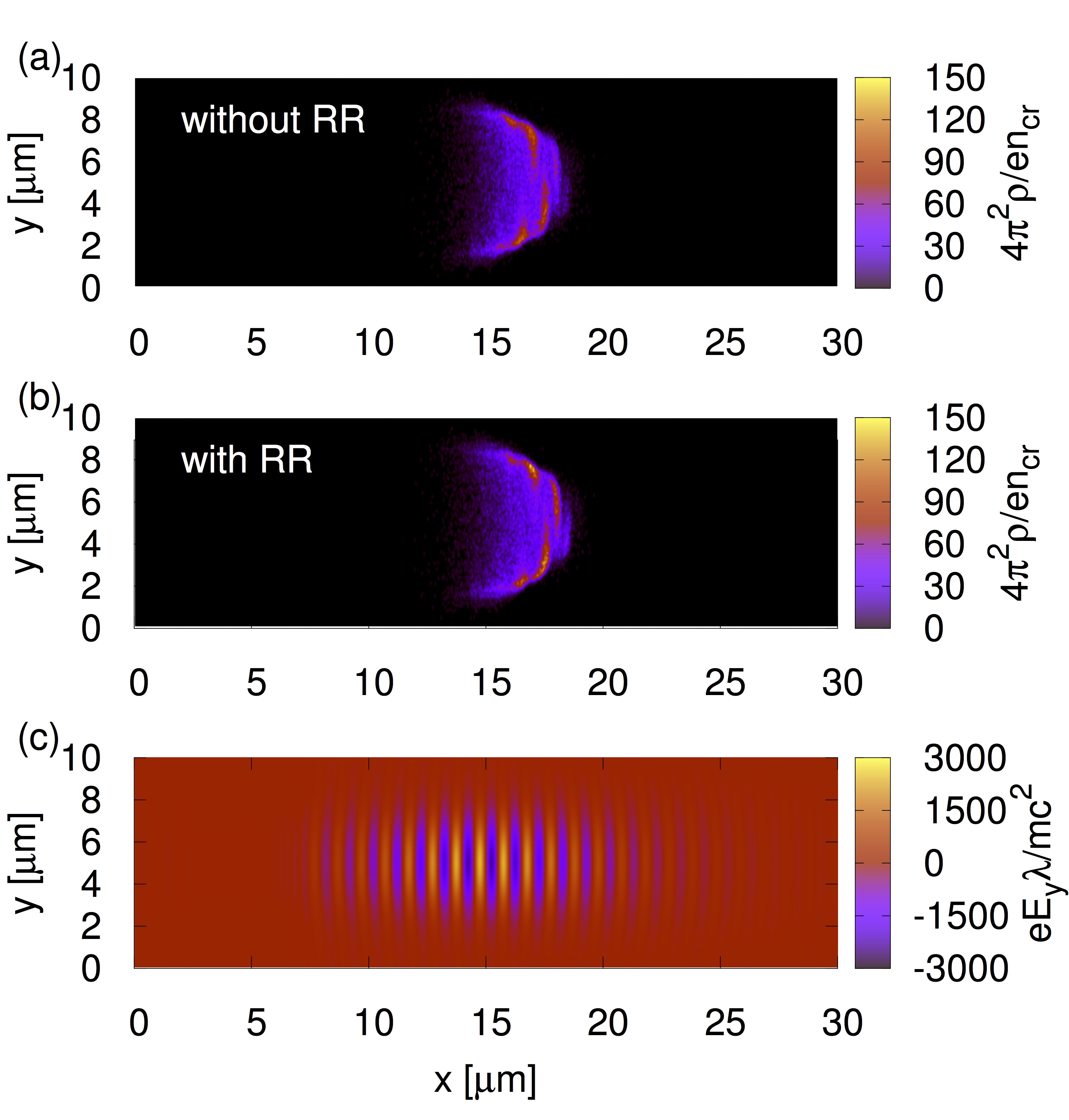}
\caption{The charge density distribution of 40 MeV electron beam for the case (a) without RR, (b) with RR, and (c) the field component $E'_y$ at the laser intensity of $2\times10^{23}$ W/cm$^2$ at the laser focusing region. }
\label{2102340MeVRRNORR600}
\end{center}
\end{figure}

\noindent When speaking in term of the classical radiation-dominant regime, the cumulative effects of RR are strong at the laser intensity of $10^{23}$ W/cm$^2$ such that $R_CN>1$. However, the situation is reversed if we assumed a tightly focused laser pulse. This is due to the utilization of focusing laser that leads to the strong ponderomotive force. The ponderomotive force repels the electrons to a region of diminishing field strength and prevents the electrons from entering the laser pulse. At the edge of the laser beam, the field strength is certainly weak and therefore the effects of RR is small. This is the case of $\kappa\lesssim 1$. For instance, a laser field with the wavelength of 1 $\mu$m focused to a waist radius of 2 $\mu$m with $\gamma\sim 80$ gives $f_{RR}/f_{P}\sim 0.04$. This situation is illustrated in Figs. \ref{2102340MeVRRNORR600} (a) $\&$  \ref{2102340MeVRRNORR600} (b) while Fig. \ref{2102340MeVRRNORR600} (c) is the corresponding laser pulse with the field component $E'_y$. Obviously, the charge density of the electron beam for both cases do not show a significant difference for the case with and without RR.
We then examine the effects of RR through the emission spectra and the photon number distribution.

\begin{figure}[h]
\centering
\includegraphics[scale=0.28]{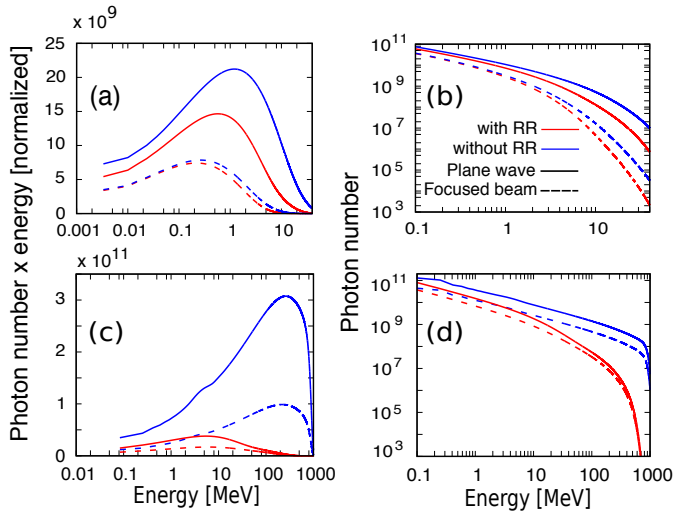}
\caption{The comparison of (a) $\&$ (c) emission spectra, and (b) $\&$ (c) photon number distribution for the case with RR (red lines) and without RR (blue lines) at the laser intensity of $2\times10^{23}$ W/cm$^2$ in a plane wave (solid lines) and in a focused beam (dashed lines). The initial kinetic energy of electron beam is 40 MeV in (a) $\&$ (b) while 1 GeV in (c) $\&$ (d).}
\label{SpecIRRNORR}
\end{figure}

The comparison of emission spectra and photon number distributions are shown in Fig. \ref{SpecIRRNORR} at the laser intensity of $2\times10^{23}$ W/cm$^2$ for the case of RR and without RR in a plane wave with temporal profile and in a focused beam. In Fig. \ref{SpecIRRNORR} (a), we observed that not only the emission spectra are strongly suppressed but the difference between the case of RR and without RR become small when a focused beam is considered as compared to the plane wave case. Meanwhile, the suppression of photon production is also noted as shown in Fig. \ref{SpecIRRNORR} (b). For the case of the focused beam, the electrons emit about $1.5 \times 10^{18}$ photons per pulse in 0.1 $\%$ bandwidth at the photon energy of 15 MeV for the case with RR, i.e. one order smaller than the case without RR. However, the photon production rate is many order larger if one only consider the plane wave field. The suppression of the effects of RR can be easily avoided by carefully choosing the energy of the incoming particle such that $\kappa>1$ and then resulting in $f_{RR}/f_{P}>>1$. For instance, when the electron beam with the energy of 1 GeV is considered the difference between the case with RR and without RR becomes clear for the focused beam as well as the plane wave as shown in Figs. \ref{SpecIRRNORR} (c) $\&$ (d).  

\begin{figure}[h]
\begin{center}
\includegraphics[scale=0.28]{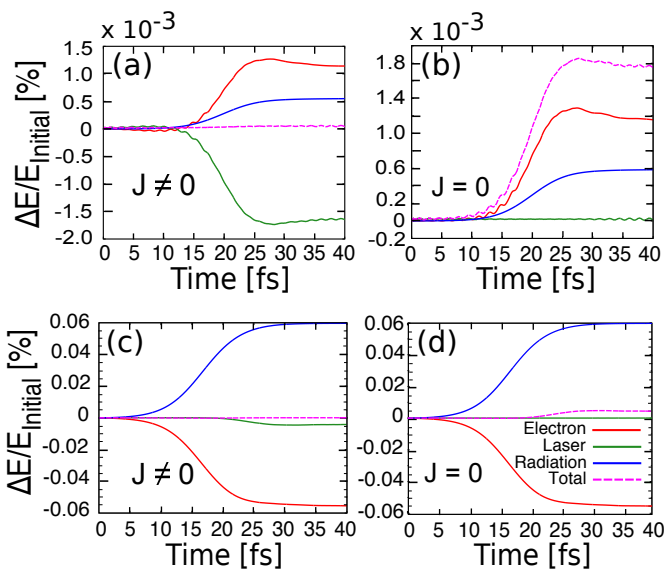}
\caption{The time evolution of energy conservation (a) with current feedback, and (b) without current feedback at the laser intensity of $2\times10^{23}$ W/cm$^2$ for 40 MeV electron beam. The time evolution of energy conservation at the same laser intensity for 1 GeV electron beam (c) with current feedback, and (d) without current feedback. The solid lines indicate each term of Eq. \ref{Etotal} while the dashed line being the total energy balance.}
\label{ENE_2102340MeVconvs}
\end{center}
\end{figure}

In addition to the suppression of RR, the accompanied process is the depletion of the laser field energy. The time evolution of energy balance with and without current feedback with the effects of RR at the laser intensity of $2\times10^{23}$ W/cm$^2$ are shown in Fig. \ref{ENE_2102340MeVconvs}. The solid lines indicate each term of Eq. \ref{Etotal} while the dashed line being the total energy balance. If the total energy is conserved, the total energy balance should be zero throughout the simulation. 

However, the loss of laser field energy may and may not be converted into radiation emission. For example, in Fig. \ref{ENE_2102340MeVconvs} (a), the laser field energy depletion is clear when the current feedback is included. This is the case where $\kappa\lesssim 1$ and $f_{RR}/f_P<<1$ and the major force that contributes to the work is the ponderomotive force. Thus, the depletes of laser field energy is accompanied by the increases of electron effective mass in a similar amount, i.e. $\Delta\mathcal{E}_{field}=\Delta\mathcal{E}_{e}$. Then, part of the energy gain is emitted into radiation emission. Therefore, the energy of the laser is transferred to the electrons and then a fraction of this energy is converted into radiation emission. There is about 4 mJ of energy lost from laser and 2.5 mJ of it is being absorbed by the electrons and the rest is converted into radiation emission when the current feedback is included as shown in Fig. \ref{ENE_2102340MeVconvs} (a). In this case, there is only 37.5 $\%$ of energy losses from the laser is converted into gamma-rays. On the contrary, there is about 4.6 mJ of energy missing in the case without current feedback as shown in Fig. \ref{ENE_2102340MeVconvs} (b).  

The situation is reversed when the electron beam with the energy of 1 GeV is considered as shown in Fig. \ref{ENE_2102340MeVconvs} (c). In this case, $\kappa>1$ and $f_{RR}/f_P>>1$. The electrons are able to penetrate into the stronger field region. As a consequence, the electrons lost most of their energies into radiation emission via RR. At the same time, the energies absorbed by the electrons from the laser field are also fully converted into radiation emission. Although the laser field depletion is small but it is essential to ensure the conservation of the total energy throughout the interaction as shown in Fig. \ref{ENE_2102340MeVconvs} (d) in which the energy is not conserved when the current feedback is neglected.

From the observations, we note that the energy of radiation emission in laser-electron beam collision comes from electron beam and only a small amount contribution from the laser. It is therefore not sufficient to take $\Delta\mathcal{E}_{rad}/\mathcal{E}_{laser}$ as the conversion efficiency of laser energy into radiation emission when the effects of RR is being considered. Then the expression $\Delta\mathcal{E}_{rad}/(\Delta\mathcal{E}_{field}+\mathcal{E}_{e})$ is more consistent for energy conversion efficiency. This can be read as the radiation energy conversion from the initial kinetic energy of the electron beam plus the electromagnetic work done by the laser on the electrons. For the case in Fig. \ref{ENE_2102340MeVconvs} (a) $\&$ (b), the energy conversion is 14 $\%$ with current feedback while 23 $\%$ without current feedback respectively. When an electron beam with energy of 1 GeV is used, it converts up to 89 $\%$ of the total initial energy into radiation emission.

\section{Summary}
In summary, we have shown the 3D PIC simulation results for the interaction of a pulsed Gaussian beam at laser intensity $2\times10^{23}$ W/cm$^2 $ with 40 MeV and 1 GeV electron beam. We observed that the ponderomotive force plays a crucial role in determining the radiation emission at the laser intensity $\sim10^{23}$ W/cm$^2$. This force serves as a barrier to the electron with lower energy to enter the laser pulse, and resulting in the reduction of the radiation emission followed by the laser field depletion. Also, the acceleration in the direction of EM wave propagation is not effective for radiation and its reaction. This is obvious in the formulation of the quantum parameter $\chi$ and emission cross section. 

Although the laser energy depletion does not significantly alter the radiation process, however, the advantage of self-consistent study is being to understand the energy breakdown of a radiative process. This may further give a significant impact for the studies of electron-positron pairs production especially if the electron beam charge of a few nano Coulomb is considered.


To generate high-intensity gamma ray with high conversion efficiency at the energy range of 10 - 20 MeV by using the high-intensity laser above $10^{23}$ W/cm$^2 $, we emphasized that the condition $f_{RR}/f_P>>1$ is required to overcome the ponderomotive potential from the laser field. The same requirement is also necessary to observe a clear signature of radiation reaction in the radiation spectrum.

\begin{acknowledgments}
This work was performed at Large-Scale Computer System at Cybermedia Center Osaka University and Hypnos cluster at Helmholtz-Zentrum Dresden-Rossendorf (HZDR).
\end{acknowledgments}

%

\end{document}